\documentclass{aastex631}
\usepackage{hyperref} 
\usepackage{natbib}
\usepackage{textcomp}
\usepackage{gensymb}
\usepackage{xcolor}
\usepackage{amsmath}
\usepackage{comment}
\usepackage{graphics}
\usepackage{array}
\usepackage{tabularx}
\usepackage{booktabs}
\usepackage{amsmath,amssymb}
\usepackage{tablefootnote}
\usepackage[utf8]{inputenc}
\usepackage{newunicodechar}

\newunicodechar{ʼ}{'}

\begin{document}
\title{The disrupted chemical enrichment history of the Milky Way driven by gas accretion}
\author[0000-0001-5258-1466]{Jianhui Lian}
\affiliation{South-Western Institute for Astronomy Research, Yunnan University, Kunming, Yunnan 650091, People’s Republic of China}
\affiliation{Yunnan Key Laboratory of Survey Science, Yunnan University, Kunming, Yunnan 650500, Peopleʼs Republic of China}
\email{jianhui.lian@ynu.edu.cn (JL)}  

\author{Qinhao Shao}
\affiliation{South-Western Institute for Astronomy Research, Yunnan University, Kunming, Yunnan 650091, People’s Republic of China}

\author{Min Du}
\affiliation{Department of Astronomy, Xiamen University, Xiamen, Fujian 361005, China}
\email{dumin@xmu.edu.cn (DM)}

\author{Hanyuan Zhang}
\affiliation{Institute of Astronomy, University of Cambridge, Madingley Road, Cambridge CB3 0HA, UK}

\author{Xinxin Duan}
\affiliation{Department of Astronomy, Xiamen University, Xiamen, Fujian 361005, China}

\author{Xin Wang}
\affiliation{School of Astronomy and Space Science, University of Chinese Academy of Sciences, Beijing 100049, China}
\affiliation{National Astronomical Observatories, Chinese Academy of Sciences, Beijing 100101, China}
\affiliation{Institute for Frontiers in Astronomy and Astrophysics, Beijing Normal University,  Beijing 102206, China}

\author{Zheng Zheng}
\affiliation{National Astronomical Observatories, Chinese Academy of Sciences, Beijing 100101, China}
\affiliation{Key Laboratory of Radio Astronomy and Technology, Chinese Academy of Sciences, Beijing, 100101, China}


\begin{abstract}
As the only galaxy enabling temporally-resolved observations from an internal vantage point, the Milky Way as a galaxy offers unique insights into galactic chemical enrichment history, {establishment of} fundamental scaling relations, and {the} underlying astrophysical processes. However, this insider perspective also introduces strong selection effects, hindering direct measurement of the Milky Way's global properties and comparison with the broader galaxy population, for the vast majority of which only integrated properties can be measured. {Here we report our measurements of the Milky Way's temporally-resolved galaxy-scale average metallicity using data from the APOGEE survey after correction for the selection function.} Our findings unveil a present-day metallicity of the Milky Way close to the Sun, { an interrupted} integrated age-metallicity relation, and {a disturbed} evolutionary trajectory in the mass-metallicity diagram, likely caused by dilution and inside-out growth associated with external gas accretion around 7~Gyr ago. 
Our results highlight the critical role of gas accretion in disrupting the galactic enrichment histories and introducing scatter in mass-metallicity relations.
\end{abstract}

\section{Introduction}
Fundamental scaling relations between heavy element abundance (i.e., metallicity) in gas and stars versus the stellar mass of galaxies has been established for decades \citep{lequeux1979,Tremonti2004,Gallazzi2005}. These scaling relations are found to evolve significantly from the early Universe at $z>3$ to the present day \citep{maiolino2008,sanders2021}. Although the mass-metallicity relations have been found for a long time, their origin remains unclear. Various scenarios have been proposed, generally invoking a mass-dependence in one or some galactic properties, including outflow strength \citep{larson1974,Tremonti2004,lian2018a}, star formation efficiency \citep{brooks2007}, gas accretion-induced dilution \citep{dalcanton2004}, and initial mass function \citep{koppen2007,lian2018a}. This variety of explanations reflects degeneracies between the parameters that regulate chemical enrichment in galaxies. A critical limitation in analyzing extra-galactic observations is the reliance on statistical analysis that assumes coherent evolutionary histories of galaxies across diverse galaxy masses and redshifts.  

As the closest galaxy that allows detailed, temporally-resolved observations of individual stars, the Milky Way provides a unique perspective from a single galaxy's enrichment history to understand the establishment and cosmic evolution of the extra-galactic mass-metallicity relations, as well as the underlying astrophysical processes. However, unbiased measurements of the Milky Way's global chemistry properties remain elusive due to strong selection effects from our inside perspective. Our embedded position within the disk creates a complicated spatial selection function and excludes stars from the densest inner Galaxy regions where the dust extinction is high. 

Over the past decade, the prevalence of massive stellar spectroscopic surveys is making direct measurements of the Milky Way's global chemistry properties feasible. In particular, the APOGEE survey \citep{majewski2017}, which operates in the near-infrared to minimize dust extinction, has mapped a unique sample of $\sim0.6$ million stars with comprehensive radial coverage from the Galactic bulge to the outer disk \citep{zasowski2013,zasowski2017,beaton2021,santana2021}. Leveraging the wide spatial coverage of this survey, in an earlier attempt \citep{lian2023}, we obtained the integrated metallicity profile of the Milky Way within $2<R<15$~kpc, after carefully accounting for the APOGEE selection function. Here we move a step further to derive the globally integrated metallicity of our Galaxy, providing the first integrated age-metallicity relation of a single galaxy that is free of age-metallicity degeneracy, which is difficult to account for in stellar population synthesis in extra-galactic studies. 

\section{Data and Methods}
\subsection{Data}
This paper is the third in a series aimed at deriving the integral properties, specifically the globally averaged metallicity in this work, of the Milky Way galaxy. We use the data from the APOGEE stellar spectroscopic survey \citep{majewski2017}, which is currently the only completed survey that has spectroscopically mapped stars in both the bulge, inner disk, and outer disk. We adopt measurements of chemical abundances from the APOGEE catalog and ages and distances from the astroNN value-added catalog \citep{mackereth2019,leung2019} from SDSS Data Release 17 \citep{blanton2017,sdss-dr17}. We select $\sim$230,000 giant stars (log$(g)<3.5$) from the main survey (target flag {EXTRATARG}==0) with high-quality spectra (signal-to-noise$>$50).

To interpret our observational results, we investigate TNG50 cosmological hydrodynamical simulation  \citep{Pillepich2019,Nelson2019}.  
For each simulated galaxy, we recenter the system on the minimum of the gravitational potential, subtract the bulk velocity, and determine the orientation based on the stellar angular momentum, constructing both face-on and edge-on projections using \texttt{pynbody} \citep{pontzen2013}. 

\subsection{Correction for the selection function}
In order to recover the intrinsic density distribution of mono-abundance populations, we convert the density of {giant} stars observed  with APOGEE to that of the underlying population by utilizing our precise knowledge regarding stellar evolution. The procedure to derive the intrinsic 3D mass and luminosity density distribution of the Milky Way are detailed in \citet{lian2022b,lian2023}. Here we briefly summarize the main steps and refer the readers to those papers for more details. 

We first bin our observed sample and calculate the observed number density of stars in a high-dimensional space of 3D position, [Fe/H], and [Mg/Fe]. The binning scheme in mono-abundance populations (MAPs) ranges from -1.1 to 0.5~dex (with an even step of 0.2~dex) in [Fe/H] and -0.1 to 0.4~dex (with a step of 0.1~dex) in [Mg/Fe]. 

Then we generate mock catalogs of stars in this high-dimension space by sampling the state-of-the-art PARSEC isochrones \citep{bressan2012}, assuming the Kroupa initial mass function \citep{kroupa2001} and using a 3D dust extinction map \citep{bovy2016,green2019}. By applying the selection criteria of the APOGEE survey and of our sample to the mock catalog, we obtain the probability of a star being selected as targets and finally observed.  
APOGEE survey targets are selected from 2MASS catalog via (J-K$_{\rm s}$)$_0$-H color-magnitude diagram, where (J-K$_{\rm s}$)$_0$ is the extinction-corrected color \citep{zasowski2013,zasowski2017,beaton2021,santana2021}. Due to limited number of fibers, only a certain fraction of stars randomly selected from the suitable targets are observed, from which our sample is selected based on the quality criteria described above. 
Ultimately, the selection function is a combination of these two components: the probabilities been selected as targets and {then} observed and selected in the final sample. Since the target selection criteria and observation fraction vary with field, the selection function is calculated field-by-field. Finally, dividing the observed density of stars by this composite probability and multiplying the mass/luminosity-to-number ratio of the mock catalog give us the intrinsic 3D mass/luminosity density distribution of MAPs. 

We further unfold MAPs in the age dimension using the observed age distribution of MAPs, considering their variations as a function of radius and height, to derive the properties of mono-age populations. Since the spectroscopic ages of high-$\alpha$ giant stars with [Mg/Fe]$>0.2$ become unreliable at metallicity [Fe/H]$<-0.5$ due to extra-mixing \citep{shetrone2019}, we follow the strategy in \citet{lian2022b} that assumes the high-$\alpha$ populations at [Fe/H]$<-0.5$ follow the same age distribution as those at {[Fe/H]$\sim-0.5$}. The binning scheme in age is a constant step of 2~Gyr from 0 to 14~Gyr. We do not perform the analysis in the three dimensions of age, [Fe/H], and [Mg/Fe] at the beginning because the observed number of stars are relatively limited, adding a new dimension will lead to too low statistics. The magnesium abundance [Mg/H] is calculated using the relation [Mg/H] = [Fe/H] + [Mg/Fe]. 

\subsection{Chemical evolution model}
The chemical evolution models presented in this work are derived from a multi-zone GCE model (\url{https://github.com/QinhaoShao/GCE_model.git})  that updates the earlier version employed by \citet{lian2020a,lian2020b}. The models incorporate the key processes governing the chemical enrichment of galaxies, including star formation, gas inflow, and gas outflow, and include nucleosynthetic yields from core-collapse supernovae \citep{limongi2018}, type-Ia supernovae \citep{Iwamoto1999}, and asymptotic giant branch stars \citep{Cristallo2015}. {In particular, we adjust the Mg yield in the core-collapse supernova yield table by increasing it by 0.3 dex \citep{Prantzos2018, lian2020c, Jost2025}.} {In the model, the star formation rate is assumed to follow the observed star formation law \citep{kennicutt1998}, scaling with the gas surface mass density to the power of 1.4; the proportionality coefficient is treated as a free parameter that sets the star formation efficiency.}

{Building on our previous GCE models that quantitatively reproduce the double [$\alpha$/Fe]--[Fe/H] sequences \citep{lian2020a,lian2020b}, we adopt a gas accretion history consisting of four phases, together with the corresponding star formation history. The initial gas-infall phase lasts for 2.5~Gyr, during which the accretion rate is assumed to be constant. This is followed by an extended phase of secular evolution, in which the accretion rate declines exponentially with an e-folding timescale of 0.5~Gyr. To reproduce both the observed double [$\alpha$/Fe]--[Fe/H] sequences and the broken age--metallicity relation \citep{lian2020b}, we introduce an additional accretion episode that begins at a lookback time of 6.3~Gyr, with a constant accretion rate before it is subsequently truncated. For this episode, the metallicity of the infalling gas is set to $-0.65$. This value is chosen to ensure that the predicted ratio of young to old stellar populations is consistent with observational constraints. A lower infall metallicity would require a less efficient accretion episode to match the chemical evolutionary tracks, thereby leading to an underproduction of young stellar populations. The late accretion episode is required to occur later than the second infall adopted in classical ``two-infall'' models, namely at a lookback time of 6.3~Gyr rather than 9.4~Gyr \citep{spitoni2019,spitoni2020}.}

An important update of the model is inclusion of the mixing effect by imposing {an analytical} description of the radial migration process assuming Gaussian diffusion of the orbit radius of stars \citep{sanders2015,frankel2018,sharma2021} as
\begin{equation}
P(R|R_0, t) = C \exp \left( -\frac{(R - R_0 + \sigma^2 t/2 R_d t_0)^2}{2\sigma^2 t / t_0 } \right), 
\label{eqn:migration}
\end{equation}
where R$_0$ denotes the birth radius of a star and $R_d$ represents the scale length of mono-age populations, for which we use the best-fitted results from \citep{zhang2025b}. $\sigma$ indicates the width of Gaussian redistribution function at a lookback time of $t_0$ that is set to be 8~Gyr, a quantity tracing the strength of the stellar radial migration in units of ${\rm kpc}$. Here we adopt the best-fitted $\sigma$ of 3~${\rm kpc}$ in \citet{frankel2020}. The constant $C$ is a normalization factor, which ensures that the integral of the probability distribution equals to one.

\begin{figure*}
	\centering
	\includegraphics[width=\textwidth]{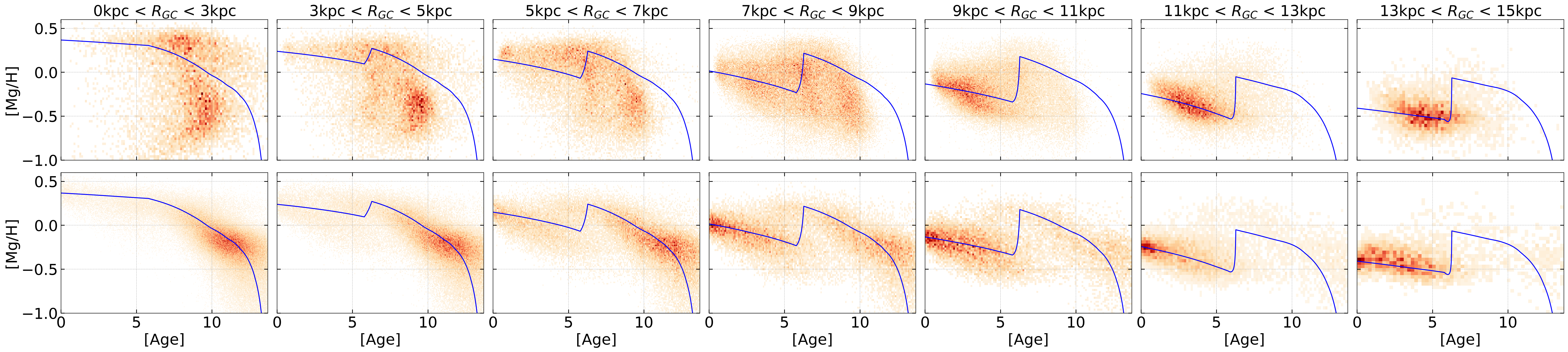}
	\caption{Observed and predicted age-metallicity distribution as a function of radius. The top row displays the observed distribution derived from APOGEE data, while the bottom row shows the corresponding predictions from the late-accretion chemical evolution model. The solid blue lines in both rows illustrate the evolutionary tracks of the late-accretion models.}
    \label{age-mgh-r}
\end{figure*}

{To reproduce the observed variation of the age-metallicity distribution with Galactocentric radius, we vary the model parameters that regulate gas accretion and star formation, including the gas accretion rates during the initial phase ($A_{\rm initial}$) and second accretion phase ($A_{\rm second}$), and the coefficient of the star-formation law for the four phases of the star formation history ($C_{\rm SFL,initial}$, $C_{\rm SFL,secular}$, $C_{\rm SFL,second}$, $C_{\rm SFL,post}$). The adopted values of these parameters for our late-accretion model are listed in Table~\ref{tab:pa}, and the resulting age–metallicity distribution at different Galactic radii are displayed in Figure~\ref{age-mgh-r}.}

{Recently, \citet{Dubay2026} demonstrated that the two-infall chemical evolution models fail to reproduce the high metallicities observed in intermediate-age stars. We argue that this discrepancy stems from the overly early onset of the second infall phase, which causes early dilution and consequently lowers the metallicities of intermediate-age stars. Adopting a later onset for the second accretion episode, as in our late-accretion model, mitigates this discrepancy as shown in Figure~\ref{age-mgh-r}.}

\begin{table}
\centering
\caption{Adopted parameters for the late-accretion GCE model.}
\label{tab:pa}
\resizebox{\textwidth}{!}{
  \begin{tabular}{cccc *{6}{c}}
    \hline
    \hline
    R & A$_{initial}^a$ & A$_{second}$ & C$_{\mathrm{SFL,initial}}$ & C$_{\mathrm{SFL,secular}}$ & C$_{\mathrm{SFL,second}}$ & C$_{\mathrm{SFL,post}}$ \\
    (kpc) & (M$_{\odot}$yr$^{-1}$kpc$^{-2}$) & (M$_{\odot}$yr$^{-1}$kpc$^{-2}$) & (M$_{\odot}$yr$^{-1}$kpc$^{-2}$) & (M$_{\odot}$yr$^{-1}$kpc$^{-2}$) & (M$_{\odot}$yr$^{-1}$kpc$^{-2}$) & (M$_{\odot}$yr$^{-1}$kpc$^{-2}$) \\
    \hline
0-3 & 0.13 & 0 & 1.4 & 0.5 & 0.5 & 0.15 \\
3-5 & 0.11 & 0.05 & 1.4 & 0.5 & 0.5 & 0.14 \\
5-7 & 0.07 & 0.09 & 1.5 & 0.5 & 0.5 & 0.14 \\
7-9 & 0.05 & 0.17 & 1.6 & 0.5 & 0.5 & 0.1 \\
9-11 & 0.03 & 0.22 & 1.8 & 0.5 & 0.5 & 0.06 \\
11-13 & 0.006 & 0.27 & 2 & 0.3 & 0.3 & 0.06 \\
13-15 & 0.004 & 0.32 & 2.4 & 0.3 & 0.3 & 0.02 \\
    \hline
Note $^a$: Gas accretion rate. 
  \end{tabular}
}
\end{table}

\section{Results and Discussion}
\subsection{The Milky Way's average metallicity}
\begin{figure*}
	\centering
	\includegraphics[width=12cm]{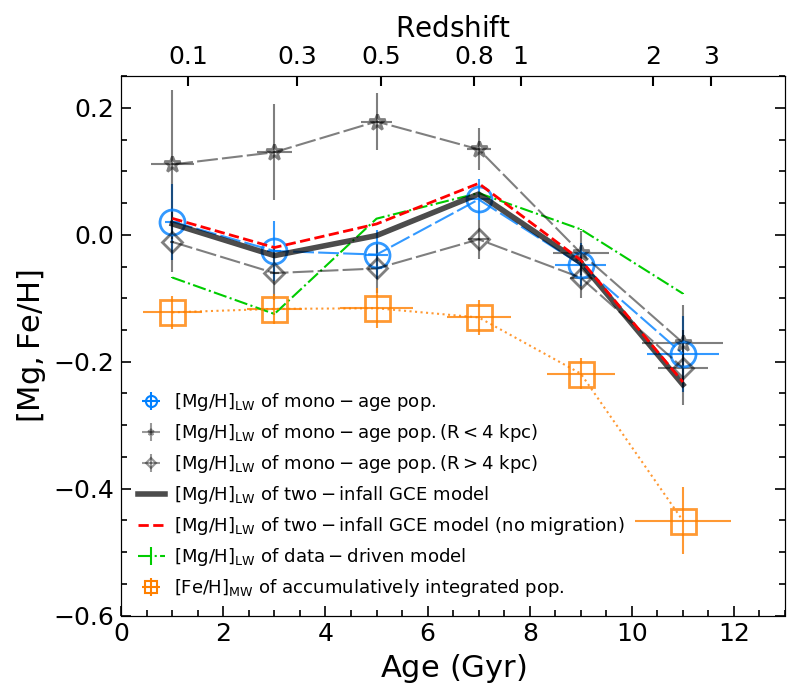}
	\caption{Average metallicity of the {Milky Way galaxy} as a function of age. The light-weighted average magnesium abundance ([Mg/H]$_{\rm LW}$) of mono-age populations, which is {integrated over the entire Galaxy and} comparable to the gas-phase metallicity measurements of external galaxies at different redshifts, are shown in blue circles. [Mg/H]$_{\rm LW}$ of the Galaxy within and beyond 4~kpc are shown in grey stars and diamonds, respectively. Black solid and green dash-dotted lines depict the prediction of chemical evolution model involving a recent gas-accretion event \citep{lian2020b} and data-driven model best-fitting the resolved age-[Fe/H] relations across the Galaxy \citep{zhang2025b}, respectively. Red dashed curve shows the {late-accretion} model with no radial migration. The mass-weighted integrated iron abundance ([Fe/H]$_{\rm MW}$) at each lookback time, which is comparable to the stellar metallicity of galaxies, are shown in orange squares. Error bars denote $1\sigma$ uncertainty of the measurements.}
    \label{feh-age}
\end{figure*}

To compare with external galaxies where only integrated measurements are available, we carefully choose two integrated chemistry properties of the Milky Way that have consistent astrophysical meanings with the gaseous and stellar metallicity of galaxies. The gas-phase metallicity of galaxies is represented by the abundance of oxygen relative to hydrogen (O/H) in the interstellar medium. The abundance of iron and $\alpha$ elements (e.g., oxygen, magnesium, silicon, etc.) in the surface of a star are generally invariant after its formation, serving as a fossil record of the element abundances of the gas from which the star was born. Since magnesium and oxygen are both $\alpha$ elements which are mainly synthesized and released by core collapse supernovae \citep{kobayashi2020}, their abundances in stars normalized to the solar abundance pattern ([Mg/H] and [O/H]) are generally consistent with each other. As magnesium is the most well measured $\alpha$ element from APOGEE spectra, we derive light-weighted average magnesium abundance ([Mg/H]$_{\rm LW}$), using optical $r$ band luminosity as weight, for mono-age populations of the Milky Way. This quantity traces the $\alpha$ element abundance of the Galaxy and is comparable to the gas-phase oxygen abundance of galaxies. 

Unlike gaseous metallicity traced by instantaneous oxygen abundance, stellar metallicity of an external galaxy represents the average metal abundance of all stars born to date. Observationally, the stellar metallicity measurements in galaxies are mainly constrained by iron absorption lines and derived assuming solar $\alpha$ abundance, thus actually represented by the integrated iron abundance. Therefore we use the mass-weighted average iron abundance ([Fe/H]$_{\rm MW}$) of entire populations of the Milky Way to compare with the mass-weighted stellar metallicity of local galaxies. We understand that there may still be inconsistency between the measurements in the Milky Way and other galaxies, given the dramatic differences in methodology in determining them. We expect that the quantitative comparison may be affected by the possible inconsistency to some extent, but the comparison in overall evolution trend should be robust.

To derive the Milky Way's integrated chemistry properties, we first integrate the intrinsic density distribution in physical space to obtain the total mass and luminosity of each mono-abundance, mono-age population, which are later used as weights to derive mass- or light-weighted average metallicity. The spatial integration is performed
within radius $R\leq14$~kpc, or $\sim$2.5 effective radii \citep{lian2024b}, and height $|Z|<4$~kpc. The measurements of the galaxy-scale integrated metallicities of the Milky Way are presented in Figure~\ref{feh-age}. For the present-day Milky Way, we obtain [Mg/H]$_{\rm LW}$ = $0.021\pm$0.022(syst.)$\pm$0.060(stoch.) and [Fe/H]$_{\rm MW}$ = $-0.122\pm$0.022(syst.)$\pm$0.025(stoch.)~dex. These values mean that while the instantaneous gas-phase metallicity of the Milky Way has reached solar levels, the stellar metallicity averaged over our Galaxy's star formation history remains sub-solar. The offset between these two measurements is consistent with difference between gaseous and stellar metallicity of similar mass star-forming galaxies in the low-redshift Universe \citep{lian2018a}.

{The systematic and stochastic uncertainties of the integrated average metallicities and ages presented in this work are estimated following \citet{lian2023,lian2024b}. The stochastic uncertainties are inherited from those of the total mass (or luminosity) and the [Fe/H] (or [Mg/H]) of each mono-abundance population (MAP). To estimate the uncertainty in mass and luminosity, we perform a Monte Carlo simulation assuming Poisson uncertainties for the observed 3D number densities: we perturb each observed density within its uncertainty and recompute the mass and luminosity of each MAP over 100 trials, adopting the standard deviation of the resulting values as the stochastic uncertainty. For the [Fe/H] (or [Mg/H]) uncertainty of each MAP,  to be conservative, we assume it to be half the [Fe/H] bin width used to define the MAPs (i.e., 0.1~dex). When unfolding each MAP in the age dimension, we also incorporate the stochastic age uncertainties of individual stars (~30\%). The stochastic uncertainties in [Fe/H] and [Mg/H] for individual stars are only $\sim$0.02~dex, giving a negligible contribution to the total uncertainty budget of the globally averaged quantities, and are therefore not considered.}

{The systematic uncertainties of our measurements mainly come from the specific choice of the stellar isochrones and dust extinction map in the calculation of the selection function. Although it is challenging to quantify the systematic uncertainty, we provide a rough estimate by calculating the variations in the measurements when adopting different isochrones or extinction maps. For this purpose, we use the alternative MESA isochrones \citep{choi2016,dotter2016} and 3D extinction map towards the inner Galaxy from \citep{chen2013}. Taking the present-day measurements as an example, the variations in integrated average metallicities ([Mg/H]$_{\rm LW}$ and [Fe/H]$_{\rm MW}$) caused by the different choice of isochrones are $\sim0.022$~dex, much higher than the variations arising from using different extinction maps ($\sim$0.003~dex). The final systematic uncertainty is assumed to be the square root of the sum of the squares of variations with different isochrones and extinction maps. We note that the systematic uncertainty of the chemical abundances of individual stars, which is at a level of $\sim$0.05~dex \citep{griffith2022}, is not included in this work. Thus the derived systematic uncertainty is likely a lower limit estimate.} 

{The uncertainties in the current measurements are dominated by stochastic uncertainties inherited from the observed 3D number density at each position, owing to the relatively limited statistics. This situation will greatly improve with the much larger samples expected from next-generation stellar spectroscopic surveys (e.g., 4MOST and SDSS-V).}

\subsection{Disrupted integrated age-metallicity relation}
Regarding the instantaneous metallicity represented by [Mg/H]$_{\rm LW}$, the integrated age-metallicity relation of the Milky Way is not monotonic, but {interrupted}. 
As shown in Figure~\ref{feh-age}, the Milky Way reached peak [Mg/H]$_{\rm LW}$ of 0.056$\pm0.036$(syst.)$\pm0.032$(stoch.)~dex at age $\sim$7~Gyr ago, which is 
followed by a mild decrease of 0.09~dex until 2~Gyr later when the metallicity resume to increase. When split into inner and outer part at $R=4$~kpc, this {disrupted} shape persists in the outer part of the Milky Way, albeit being slightly weaker, while the inner Galaxy {interrupts} at a younger age of 5~Gyr. This {interruption} in the inner Galaxy is likely caused by radial migration, with the young stars migrated inward from the disk, given their low density in the bulge \citep{lian2024b}. In the solar radius and beyond, a {non-monotonic} age-metallicity relation has already been observed based on independent observations from different spectroscopic surveys \citep{feuillet2018,xiang2022,gallart2024,roberts2026}, albeit without dedicated correction for the selection function. For the accumulatively integrated stellar metallicity, the disturbed enrichment trend seen in the instantaneous metallicity is largely smoothed out, resulting in a roughly constant stellar metallicity ([Fe/H]$_{\rm MW}$) of the Milky Way since redshift {$z\sim$0.8}. 

The origin of the {interrupted integrated} age-metallicity relation remain debated: does it reflects a non-monotonic enrichment history of our Galaxy \citep{spitoni2019,lian2020b,johnson2021,ciuca2024} or result from radial migration mixing stars born at different Galactocentric radii \citep{feuillet2018,sharma2021,chen2025}? Since internal mixing effect should not alter global properties, the persistence of this {interruption} feature in mono-age populations integrated over almost the entire Galaxy unambiguously supports the scenario of non-monotonic enrichment history. We have tested that radial migration has negligible effect to the Galactic global properties using our multi-zone chemical evolution model. 
Instead, the non-monotonic enrichment could be well explained by the dilution process induced by a recent metal-poor gas accretion event \citep{buck2020,lian2020a,barry2026}, possibly related to the first peri-center passage of Sagittarius dwarf galaxy around 5~Gyr ago \citep{laporte2018}. 
Chemical evolution model involving a recent gas accretion event can well reproduce the observed {interrupted} age-[Mg/H] relation as shown in Figure~\ref{feh-age}. The model incorporating a recent gas accretion event and dilution naturally gives rise to the two age-[Fe/H] sequences observed around the solar radius and {an interrupted} integrated average age-[Fe/H] relation.
Signatures of this recent accretion event are also found in local bursty star formation history \citep{ruiz-lara2020} and fine disk structure \citep{lian2025a}.  

Rapid inside-out growth caused by external gas accretion may be another possibility. A non-monotonic age-metallicity relation is also present when integrating the data-driven model that best-fits the age-[Fe/H] distribution across the disk between 6$<R<13$~kpc after extrapolation to cover the inner Galaxy \citep{zhang2025}. Interestingly, their model does not require {an interrupted} age-metallicity relation at each radius, but resorts to rapid size growth attributed to external gas accretion, to account for the emergence of the lower metallicity stars younger than 7~Gyr. Note that their model incorporate empirical prescriptions of star formation and chemical enrichment that are independent from each other. A more physical model self-consistently considering these two processes will be useful to further explore this possibility. Additionally, more precise age measurements of stars towards the inner Galaxy will be helpful to disentangle these two possibilities of dilution and size growth, which are both associated with external gas accretion.

\subsection{The Milky Way's disturbed enrichment trajectory}
\begin{figure*}
	\centering
	\includegraphics[width=\textwidth]{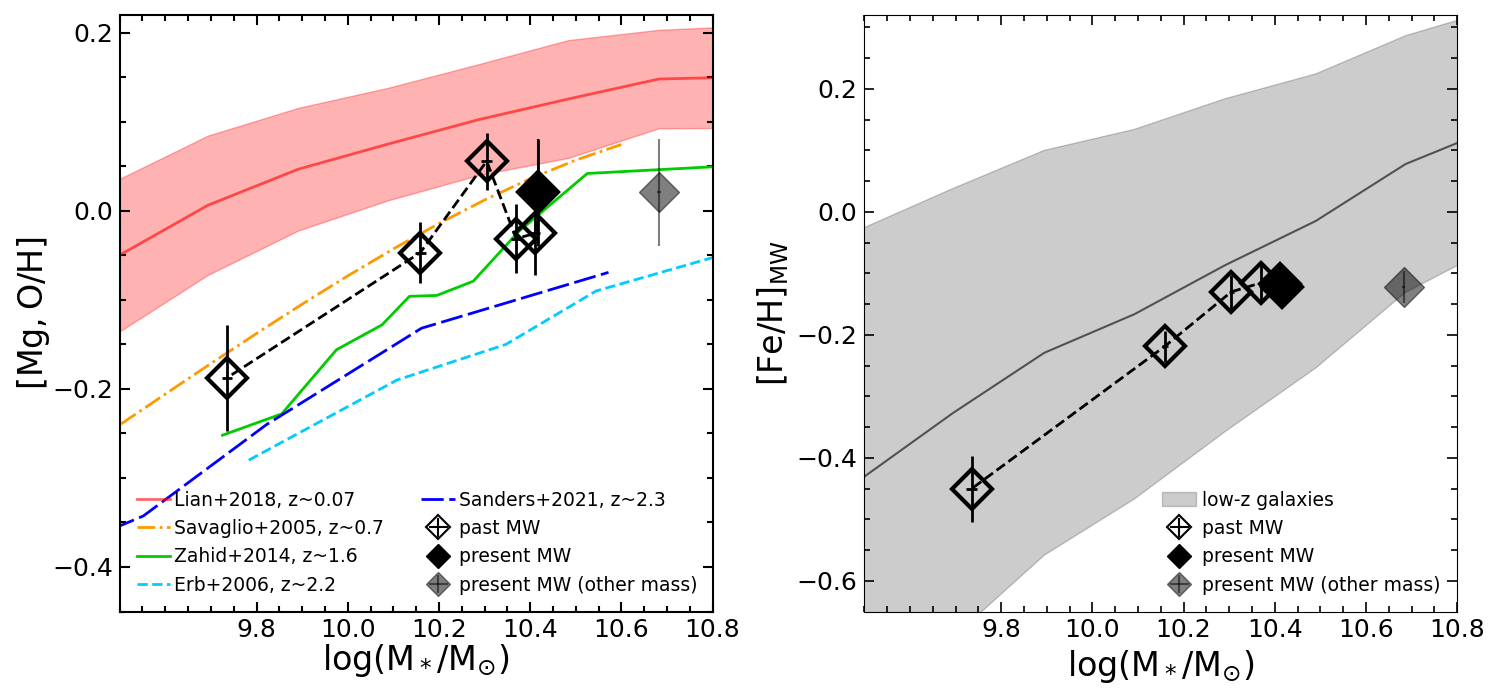}
	\caption{The Milky Way in the mass-metallicity plane. {\sl Left:} Light-weighted average magnesium abundance of the Milky Way's mono-age stellar populations in comparison with gas-phase oxygen abundance of galaxies at different redshifts. Evolution track of the Milky Way in mass-metallicity plane is presented in black diamonds.
    The measurements of the youngest age bin of 0--2~Gyr are assumed to be representative of the Milky Way at present day and indicated by filled symbols. The present-day measurement with earlier stellar mass estimate of the Milky Way that did not account for the flattened inner disk profile is shown in grey filled diamonds. All extra-galactic mass measurements have been converted to be based on Kroupa IMF \citep{kroupa2001} and oxygen abundances re-calibrated to the empirical N2 method \citep{pettini2004}. Red shaded region denotes 1$\sigma$ scatter of the distribution in low-redshift galaxies. {\sl Right:} Mass-weighted, temporally-integrated average iron abundance of the Milky Way at each lookback time in comparison with mass-weighted stellar metallicity of low-redshift galaxies. Grey solid line and shaded region indicate the median mass-metallicity relation of low-redshift star-forming galaxies presented in the left-hand panel and its 1$\sigma$ scatter, respectively.}
	\label{mass-feh}
\end{figure*}

Based on the measurements presented above, we for the first time depict the position and its evolutionary trajectory of a single galaxy, the Milky Way, in the stellar mass-gas/stellar metallicity plane. 
We adopt two sets of present-day Galactic stellar mass, a recent estimate of 2.61$\times10^{10}{\rm M_{\odot}}$ that considers a flat inner disk density profile \citep{lian2025b}, and a more widely used one of 4.81$\times10^{10}{\rm M_{\odot}}$ derived from literature mass estimates of the disk \citep{xiang2018} and bulge \citep{licquia2016}. We infer the stellar mass in the past by integrating the mass density distribution of mono-age populations that are normalized to the present-day estimate.  
For comparison, we include observational results of gas-phase metallicity in star-forming galaxies at various redshifts \citep{savaglio2005,erb2006,maiolino2008,zahid2014,lian2018a,sanders2021} and mass-weighted average stellar metallicity in low-redshift star-forming galaxies \citep{lian2018a}. The comparison is shown in Figure~\ref{mass-feh}.  

The present-day integrated chemical abundances of the Milky Way are systematically lower than the local star-forming galaxies of the same mass. This is partially explained by the interruption in the enrichment history discussed above. Another possible contribution to this inconsistency is that the stellar mass of the Milky Way may be overestimated. Previous estimates of disk stellar mass have been relying on an assumption of a single exponential disk profile. However, recent studies have reported a disk profile with a flattened inner disk component \citep{mackereth2017,lian2024b,wu2026}, suggesting a lower stellar mass of the Galactic disk. By adopting a recent {stellar mass} estimate that have taken this inner flat profile into account, the Milky Way is positioned closer to the location of local galaxies in the mass-metallicity plane. 
 
The evolution trajectories of the Milky Way in the mass-gas/stellar metallicity plane are {disturbed}, in particular in gas metallicity. The temporal decline of the Milky Way's average metallicity around 7~Gyr ago causes a notable deviation from the original evolution track in the mass-metallicity plane, providing solid evidence for gas accretion as an important source of scatter in the extra-galactic mass-metallicity relation.
The Milky Way's early evolution trajectory in the mass-metallicity plane largely follow the scaling relations observed in the distant Universe, albeit with a slightly steeper slope prior to the interruption. 
This indicates that the extra-galactic mass-metallicity relation arises naturally from the mass growth and chemical enrichment of individual galaxies. The mass dependence of galactic properties, such as star formation efficiency, outflow strength, or initial mass function, likely plays a secondary role in shaping the mass-metallicity relation. 

\subsection{Broken age-metallicity relation in the IllustrisTNG-50 simulation}
\begin{figure*}
	\centering
    \includegraphics[width=16cm]{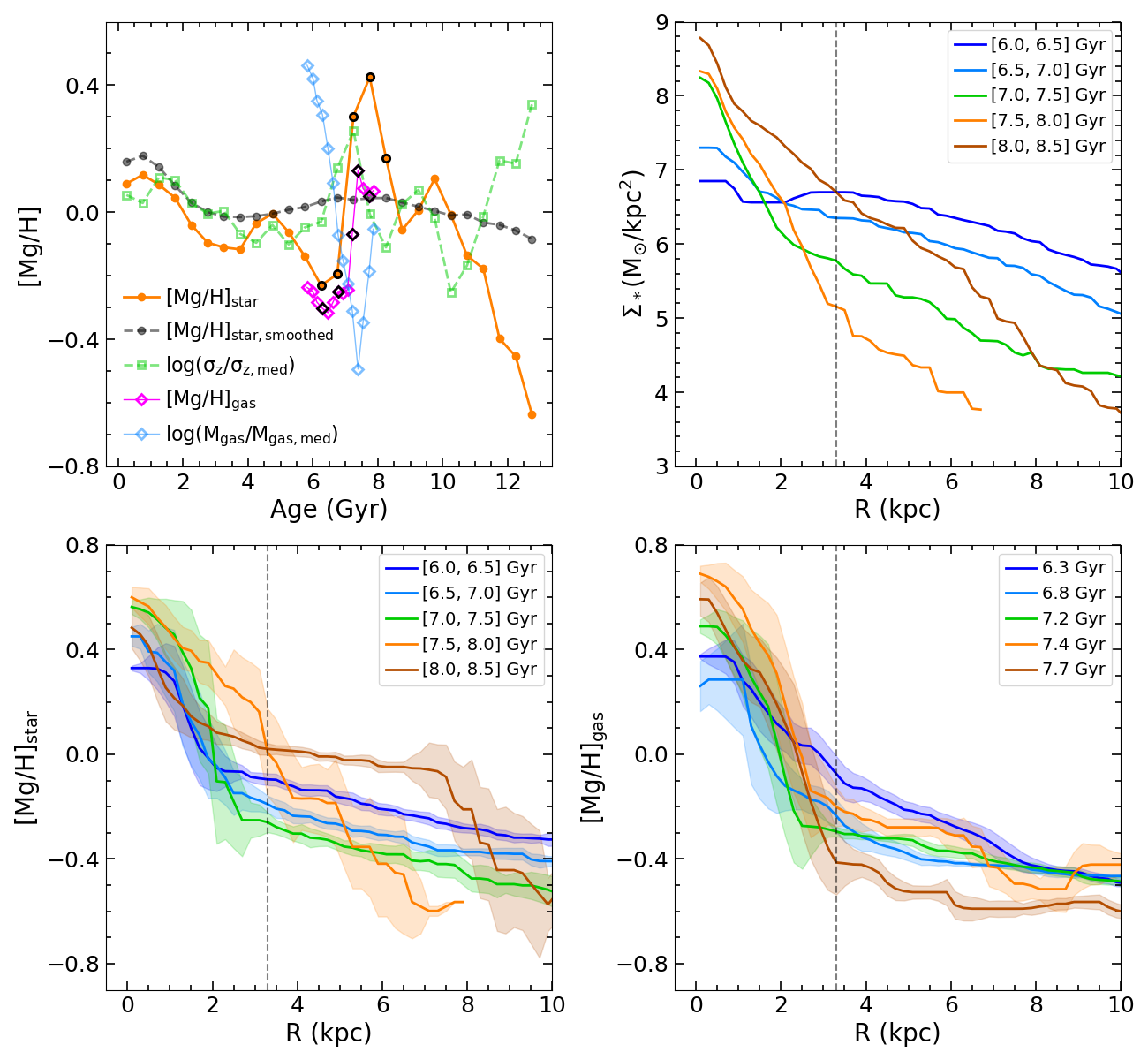}
	\caption{A case of TNG50 simulated galaxies with strong age-metallicity break. {\sl Top-left:} Temporal evolution of average metallicity of stellar and gas particles, relative vertical velocity dispersion at birth, gas mass, as shown in orange circles, purple diamonds, green squares, and blue diamonds, respectively. A smooothed age-metallicity relation considering 30\% age uncertainty of each stellar particle is presented in grey circles. Black circles and diamonds denote the five mono-age bins or snapshots for further examinations. {\sl Top-right:} Surface mass density radial profiles of five mono-age populations of stellar particles. Vertical dashed line marks the half-light radius of 3.3~kpc for this galaxy. {\sl Bottom-left:} Radial profile of light-weighted average [Mg/H] of the five mono-age populations of stellar particles. Shaded region indicate 1$\sigma$ scatter of the radial profiles.  {\sl Bottom-right:} Radial profile of mass-weighted average [Mg/H] of the gas particles in the five snapshots around the age-metallicity break. The lookback time of each snapshot is included in the legend. }
	\label{tng-case-prof}
\end{figure*}

\begin{figure*}
	\centering
	\includegraphics[width=\textwidth]{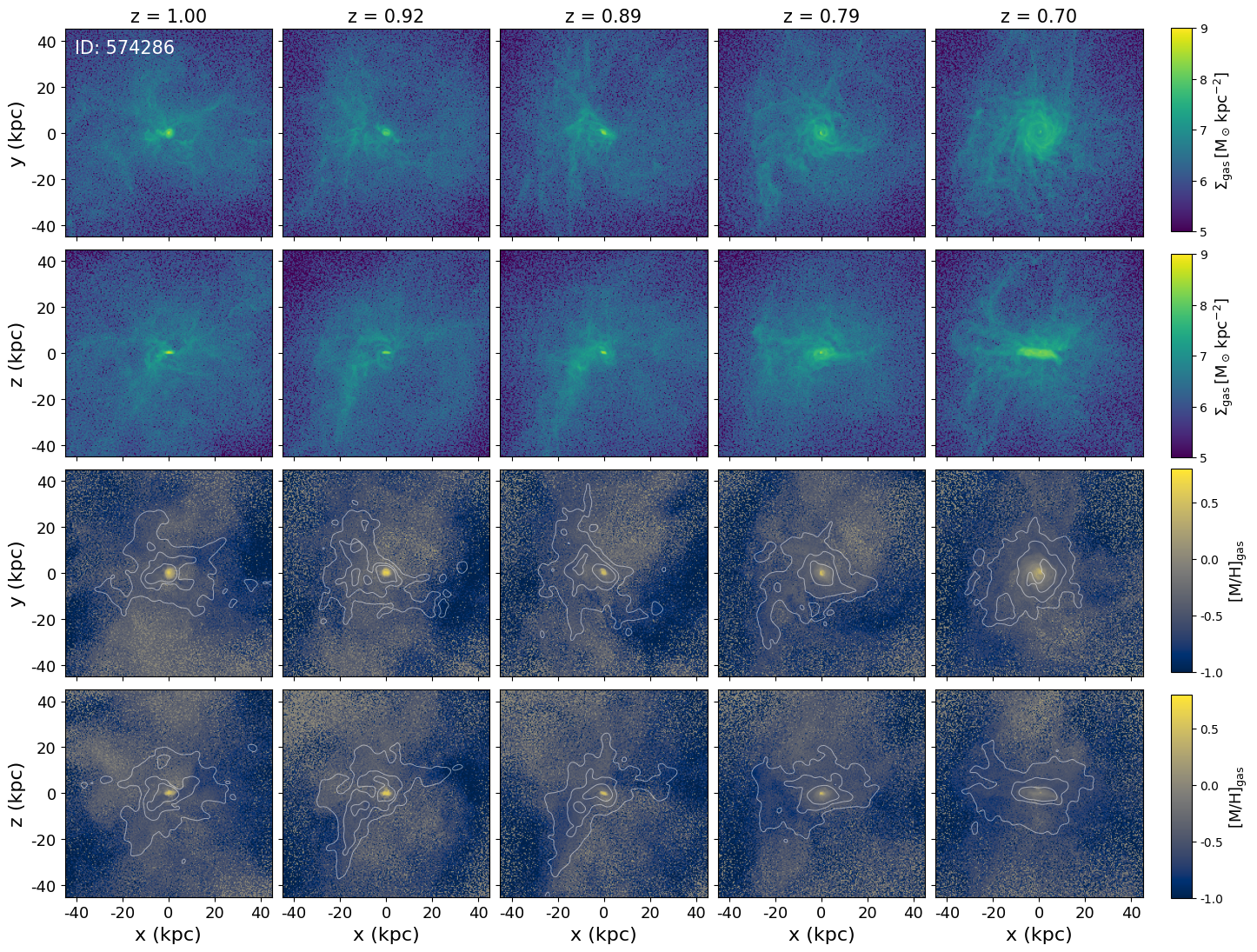}
	\caption{Density and metallicity maps of gas particles of a TNG50 galaxy with halo ID 574286. Top two rows present the density maps in X-Y and X-Z planes in five snapshots marked in Figure~\ref{tng-case-prof}. Bottom two rows are for the metallicity maps. White contours layout the density maps shown above. Each column is for one snapshot, with the corresponding redshift marked in the top.}
	\label{tng-case-map}
\end{figure*}

In addition to phenomenological models, we also explore cosmological simulations to infer the physical origin of the {interrupted} integrated age-metallicity relation. 
We employ the TNG50-1 simulation (hereafter TNG50) from the IllustrisTNG project \citep{dylan2018,jill2018,springel2018,pillepich2018b,Pillepich2019,marinacci2018}. This configuration provides both high statistical power through its galaxy population and zoom-in-equivalent resolution in $\sim$50 cMpc box. The simulation evolves $2 \times 2160^3$ resolution elements, featuring a baryonic mass resolution of $8.5 \times 10^{4}  M_{\odot}$, stellar gravitational softening length of 0.3 kpc (physical), and minimum gas softening length of 74 pc (comoving) at $z=0$.

We analyze age-metallicity distributions of stellar particles in the TNG50 simulation for massive (${\rm M_*}>10^{10} {\rm M_{\odot}}$), star-forming galaxies (specific star formation rate ${\rm >10^{-11} yr^{-1}}$) at $z=0$. To be consistent with the Milky Way's observations presented here, we calculate $r$ band luminosity-weighted magnesium abundance of each mono-age population of stellar particles. 
Notably, discontinuous age-metallicity relations emerge as a ubiquitous feature, with many systems exhibiting multiphase enrichment histories marked by several abrupt transitions. 
These sharp metallicity breaks – typically $\Delta$[Mg/H] $> 0.2$ dex within 0.5 Gyr – point to episodic dilution events. Our analysis associates these discontinuities with accretion-driven dilution events where metal-poor circumgalactic gas ([M/H]$<$-0.5~dex) or dwarf satellite material disrupts in-situ enrichment.

To explore the physical origin of the break in age-metallicity relation, we conduct a detailed analysis for one example galaxy with a pronounced break (halo ID: 547286) as shown in Figure~\ref{tng-case-prof}. While the original break is dramatic with metallicity decline of $\sim0.5$~dex, its amplitude shrinks significantly to the level observed in Milky Way after considering an age uncertainty of 30\%, which is typical for the age measurements adopted here \citep{leung2019}. To examine the behavior of gas particles, we analyzed a series of snapshots around the epoch when the age-metallicity relation is interrupted, i.e., 7~Gyr ago for this galaxy. Since gas particles themselves do not emit light and elemental abundances are not available for all snapshots examined, we calculate the mass-weighted abundance of total metals ([M/H]) as their average metallicity. 

Consistent with the picture of metal-poor gas accretion, the average metallicity of gas particles declines simultaneously with average stellar metallicity of mono-age populations by $\sim0.4$~dex. During the same time, the total gas mass increases by a factor of about three. This increase of gas mass continues after the break when both the gas and stellar average metallicity stop the rapid decline. This suggests that the dilution process is quickly balanced by the enrichment of star formation which is further enhanced due to increased gas density. 

The radial profiles of average metallicity of mono-age stellar and gas particles in the middle and right panels of Figure~\ref{tng-case-prof} further illustrate where the gas accretion and dilution occurs in this galaxy. For stellar particles, we extract five mono-age bins from 6 to 8.5~Gyr with even age interval of 0.5~Gyr, which are marked as black circles in the top left panel. The initial decline in average metallicity from 7.5--8 to 7.5--7.0 age bins is driven by systematic decline of metallicity across a wide radial range, consistent with dilution induced by gas accretion.  
However, from 7.0--7.5 to 6.5--7.0 age bins when the decline in average metallicity is more pronounced, the radial profile of stellar metallicity only drops slightly by 0.1~dex within 2~kpc and even increases beyond 2~kpc. The decline {in average metallicity } is largely due to a dramatic shift of star formation from the center to the extended disk which is more metal-poor. 
A consistent trend is also seen in gas metallicity radial profile. The decrease in globally averaged metallicity in gas particles are more pronounced than that in metallicity radial profiles. This reflects preferential gas accretion onto the outer disk, the exact cause of the radial shift of star formation. Our analysis demonstrates that the break in age-metallicity relation of this simulated galaxy is caused by both direct dilution and radial expansion of star formation, with the latter playing an more important role. Both mechanisms are results of an abrupt metal-poor gas accretion {event}.   

To identify the origin of the metal-poor gas accretion, we examine the large-scale spatial distribution of gas particles around the target galaxy within 50~kpc in the same five snapshots as Figure~\ref{tng-case-prof}. The density and metallicity maps in these snapshots are shown in Figure~\ref{tng-case-map}. The average metallicity of gas particles drops in the second and third snapshots, when a clear low-metallicity gas bridge is present in the bottom left corner of the galaxy in X-Z plane. No clear galaxy companion are associated with this gas bridge. We suspect it originates from the circumgalactic medium after being disturbed by the infalling of nearby dwarf galaxies.  

\section{Summary} %
By employing a {careful} correction for the selection function, this work presents the first measurement of the galaxy-scale average metallicity of the Milky Way galaxy and its mono-age populations. 

The metal abundance of the Milky Way at the present day turns out to be rather consistent with the metallicity in our Sun. In addition, we find the {non-monotonic} age-metallicity relation persists for mono-age populations integrated over the Galaxy, suggesting a disturbed enrichment history. The decline in metallicity at intermediate ages is likely a result of a recent gas accretion event, possibly supplied by a dwarf galaxy merger, that results in either a dilution process or rapid inside-out growth, or a combination of both. {Our analysis on TNG50 simulated galaxies reinforce the gas accretion origin of the {interrupted} age-metallicity relation. }

With these integrated metallicity measurements, we posit the Milky Way's evolution track in the mass-metallicity plane. The early phase of this track aligns well with the mass-metallicity relations of external galaxies, suggesting its establishment is largely driven by the natural mass growth and enrichment process of individual galaxies. However, at $z<1$, the Milky Way deviates from its original evolution track due to the interruption caused by gas accretion, supporting gas accretion as an important source of scatter in extra-galactic mass-metallicity relations. 

Our results demonstrate the unparalleled ability of the Galactic temporally-resolved observations in unraveling the evolutionary history of galaxies and the origin of their statistical scaling relations.   
All data presented in this work are available in a public repository \url{https://doi.org/10.5281/zenodo.20587359}.

\section*{Acknowledgements} 
{We are grateful to the reviewer for their helpful comments, which improved the robustness and clarity of this paper.}
J.L. acknowledges support by National Natural Science Foundation of China (No. 12473021), National Key R\&D Program of China (No. 2024YFA1611600), Yunnan Province Science and Technology Department Grant (No. 202105AE160021 and 202005AB160002), Key Laboratory of Survey Science of Yunnan Province (No. 202449CE340002), and the Start-up Fund of Yunnan University (No. CY22623101). 

Funding for the Sloan Digital Sky Survey IV has been provided by the Alfred P. Sloan Foundation, the U.S. Department of Energy Office of Science, and the Participating Institutions. SDSS-IV acknowledges support and resources from the Center for High-Performance Computing at the University of Utah. The SDSS web site is www.sdss.org.

SDSS-IV is managed by the Astrophysical Research Consortium for the 
Participating Institutions of the SDSS Collaboration including the 
Brazilian Participation Group, the Carnegie Institution for Science, 
Carnegie Mellon University, the Chilean Participation Group, the French Participation Group, Harvard-Smithsonian Center for Astrophysics, 
Instituto de Astrof\'isica de Canarias, The Johns Hopkins University, Kavli Institute for the Physics and Mathematics of the Universe (IPMU) / 
University of Tokyo, the Korean Participation Group, Lawrence Berkeley National Laboratory, 
Leibniz Institut f\"ur Astrophysik Potsdam (AIP),  
Max-Planck-Institut f\"ur Astronomie (MPIA Heidelberg), 
Max-Planck-Institut f\"ur Astrophysik (MPA Garching), 
Max-Planck-Institut f\"ur Extraterrestrische Physik (MPE), 
National Astronomical Observatories of China, New Mexico State University, 
New York University, University of Notre Dame, 
Observat\'ario Nacional / MCTI, The Ohio State University, 
Pennsylvania State University, Shanghai Astronomical Observatory, 
United Kingdom Participation Group,
Universidad Nacional Aut\'onoma de M\'exico, University of Arizona, 
University of Colorado Boulder, University of Oxford, University of Portsmouth, 
University of Utah, University of Virginia, University of Washington, University of Wisconsin, 
Vanderbilt University, and Yale University.

\bibliographystyle{aasjournal}
\bibliography{references_jl}{}
\end{document}